\documentclass[aps,prl,reprint,superscriptaddress,amsmath,amssymb,showpacs,longbibliography]{revtex4-1}

\usepackage[utf8]{inputenc}

\usepackage{graphicx,color}
\usepackage{amsmath,amssymb,bm}

\usepackage{physics}
\usepackage{siunitx}
\usepackage{csquotes}
\usepackage{times}
\usepackage{soul}

\usepackage{float}

\usepackage{tikz} %Zeichnen mit LaTeX
\usetikzlibrary{graphs}

\definecolor{lightblue}{RGB}{204, 243, 255}
\definecolor{middleblue}{RGB}{30,100,255}

\newcommand{\changes}[1]{{\color{black}#1}}

\date{\today}                  % Activate to display a given date or no date

\begin{document}
\title{Floquet engineering topological many-body localized systems}

\author{K.~S.~C.~Decker}
\affiliation{Technische Universit\"at Braunschweig, Institut f\"ur Mathematische Physik, Mendelssohnstraße 3, 38106 Braunschweig, Germany}

\author{C.~Karrasch}
\affiliation{Technische Universit\"at Braunschweig, Institut f\"ur Mathematische Physik, Mendelssohnstraße 3, 38106 Braunschweig, Germany}

\author{J.~Eisert}
\affiliation{Dahlem Center for Complex Quantum Systems and Fachbereich Physik, Freie Universit\"at Berlin, 14195 Berlin, Germany}

\author{D.~M.~Kennes}
\affiliation{Institut f\"ur Theorie der Statistischen Physik, RWTH Aachen University and JARA-Fundamentals of Future Information Technology, 52056 Aachen, Germany}
\affiliation{Max Planck Institute for the Structure and Dynamics of Matter, Center for Free-Electron Laser Science, 22761 Hamburg, Germany}

\begin{abstract}
We show how second-order Floquet engineering can be employed to realize systems in which many-body localization coexists with topological properties in a driven system. This allows one to implement and dynamically control a symmetry protected topologically ordered qubit even at high energies, overcoming the road block that the respective states cannot be prepared as ground states of nearest-neighbor Hamiltonians. Floquet engineering --- the idea that a periodically driven non-equilibrium system can effectively emulate the physics of a different Hamiltonian --- is exploited to approximate an effective three-body interaction among spins in one dimension, using time-dependent two-body interactions only. In the effective system emulated topology and disorder coexist which provides an intriguing insight into the interplay of many-body localization that defies our standard understanding of thermodynamics and into topological phases of matter, which are of fundamental and technological importance. We demonstrate explicitly how combining Floquet engineering, topology and many-body localization allows one to harvest the advantages (time-dependent control, topological protection and reduction of heating, respectively) of each of these sub-fields while protecting them from their disadvantages (heating, static control parameters and strong disorder).
\end{abstract}

\maketitle

In the long-standing quest for the practical realization of key quantum technologies such as quantum computing \cite{Roadmap}, a key goal is to fight off decoherence and to manipulate quantum systems in a controlled way \cite{Ladd2010}. Several promising concepts have been proposed within the past decade and became central research fields in the study of quantum many-body phenomena: %{\color{red}Topology -- the concept that phases of matter can be classified by geometric properties of their wave functions \cite{Ludwig_2015}
\changes{Topological phases of matter are reflected by robust degeneracies of ground states
and are signified by non-local order parameters 
\cite{Ludwig_2015,WenBook}.}
%-- provides an inherent geometric protection of quantum states at low energies. 
Many-body localization in disordered systems defies
our standard understanding of thermodynamics by breaking ergodicity and barring the system from thermalization \cite{mbl_basko,mbl_huse,mbl_exp1,mbl_rmp}. By this many-body localization can extend the aforementioned topological protection to high energies \cite{topmbl_huse,topmbl_hondhi}.
 Floquet engineering --- the idea that a periodically-driven non-equilibrium system 
 \cite{PolkovnikovReview,1408.5148}
 can effectively emulate the physics of a different Hamiltonian --- allows one to realize `toy models' in real-life systems and to establish stable real-time protocols to manipulate quantum states \cite{floquet1}.

%{\color{red} Traditionally, topology is understood for ground states. However, in a many-body localized system, the geometric protection caused by it holds not only for ground states but for the entire spectrum of the system. This is due to the fact that the same degree of freedom (i.e. the edge states) is decoupled from the rest of the system at all energies \cite{topmbldyn_bahri}.}

Most of the proposals on how topology, disorder, and Floquet-engineering can be exploited in the design of quantum devices \cite{PhysRevLett.116.250401,PhysRevB.98.064203,PhysRevB.98.174203,PhysRevA.96.022306,kuno2019manybodylocalization,li2019classification,PhysRevX.6.021013,PhysRevLett.118.115301,PhysRevX.6.041070,PhysRevA.100.023622,PhysRevLett.123.126401,PhysRevB.99.205419}, yet, either concentrate on one or two of these phenomena separately or use special models. %; e.g., two-body interactions, which have a crucial influence on the physics, are often neglected. 
Floquet-engineering has been suggested as a route to realizing effective topological systems \cite{Lindner2011,Sentef2015a,Hubener2017a,Topp2018a,Grushin2014,Claassen2017a,Kennes19}, but those studies neglect two-body interactions and thus 
ignore the fact that a generic driven system will heat up \cite{floquet_alessio}. Heating can be efficiently suppressed by many-body localization \cite{noheating1,noheating2}, but topological properties of a clean system are normally lost in the presence of strong disorder \cite{PhysRevLett.121.126803,PhysRevB.98.134507,orito2019}.
 
\begin{figure}[t]
  \includegraphics[width=\linewidth,clip]{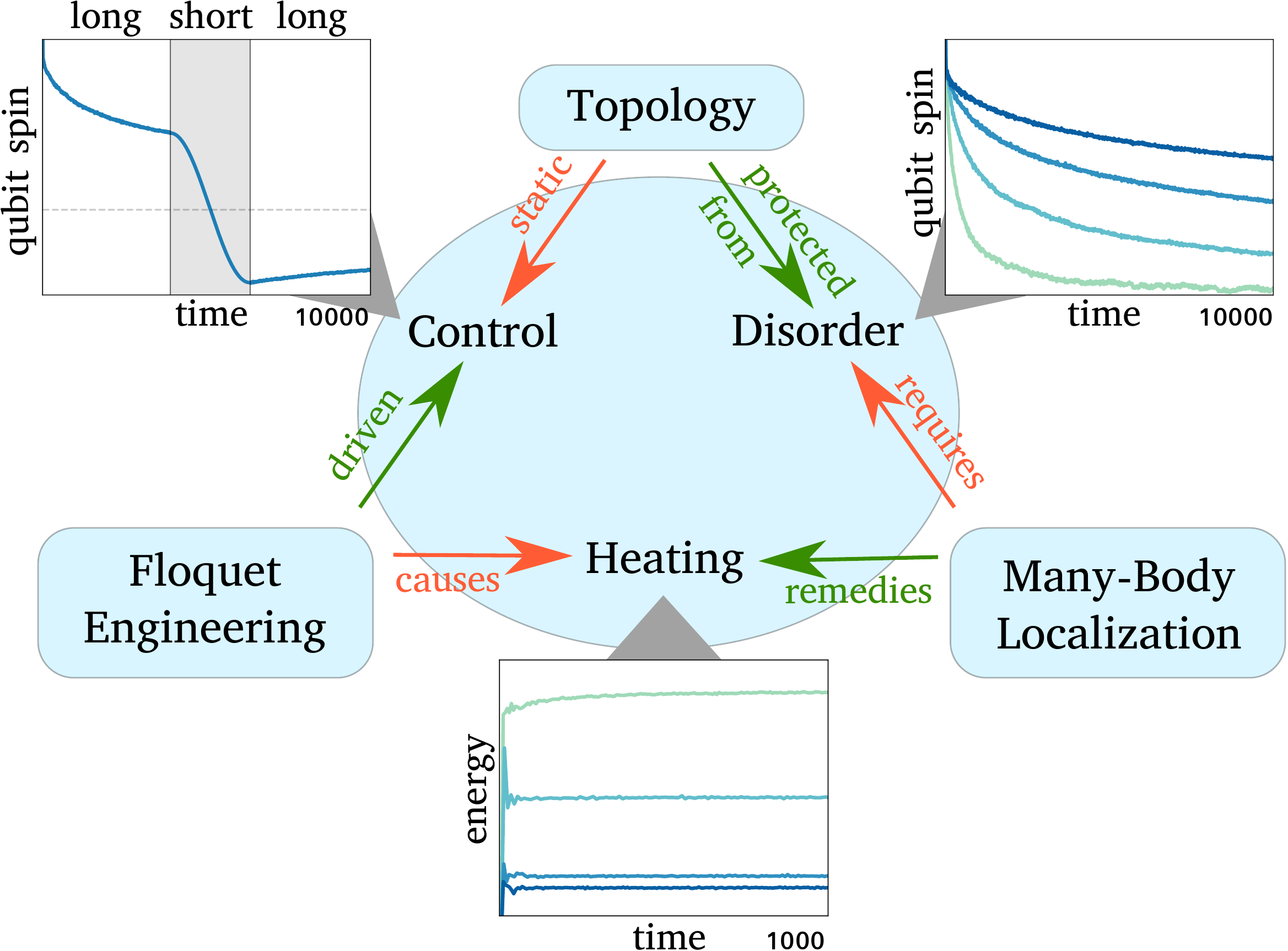}
  \caption{We combine insights from the subfields of Floquet engineering, many-body localization, and topology. 
  By doing so, we demonstrate how to harvest the full advantages promised by
these fields—flexible control, suppression of heating, and topological protection—while removing their respective disadvantages.  These advantages are accessible with programmable quantum simulators.
We explicitly illustrate this in Fig.~\ref{fig:dis} (topological protection), Fig.~\ref{fig:heating} (suppression of heating), and Fig.~\ref{fig:control} (control), which are schematically given as insets here.}\label{fig:triangle}
\end{figure}

Recently, an artificial toy model exhibiting three-body interactions was identified in which topology and many-body localization co-exist instead of hampering with each other \cite{topmbldyn_bahri,topmbldyn_yao,firstpaper,goihl2019}. Both phenomena conspire to provide topological protection of an edge spin degree of freedom even at high energies \changes{as the quantum system is localized for the {\it entire} eigen-spectrum (in contrast to only some part of it)}. 
%is localized \cite{topmbldyn_bahri,topmbldyn_yao,firstpaper}. This elevates the concept of topology and its geometric protection, traditionally understood for ground states, to the entire spectrum including arbitrary excited states or thermal ensembles thereof \cite{topmbldyn_bahri}.
 Three body-interactions \changes{as needed to realize %this 
 such artificial systems} are, however, \changes{excessively} difficult to realize in a real-life Hamiltonian, but they can naturally arise in Floquet-engineered setups, as suggested in Ref.~\cite{PhysRevLett.119.123601}. Here, we build on these ideas and extend them in four different directions: i) We show how the toy Hamiltonian of Refs.~\cite{topmbldyn_bahri,topmbldyn_yao} can be obtained by means of Floquet-engineering starting from a physical quantum spin model $t\mapsto H(t)$ with time-periodic two-body interactions. ii) We explicitly benchmark how well the effective, time-independent Floquet Hamiltonian $H^{\rm eff}$ describes the stroboscopic physics of $t\mapsto H(t)$. iii) We demonstrate to 
 what extent heating is suppressed by many-body localization. (iv) We design a dynamical protocol which allows one to flip the edge spin on a short time scale.

In sum, Floquet-engineering, many-body localization (i.e., disorder \changes{and interactions}), and topological protection are three pillars on which useful quantum applications can be built. The combination of these three ingredients elegantly counteracts all their individual shortcomings -- if one removes only one of them, one can no longer fully harvest their strengths. We now illustrate this explicitly and demonstrate the co-existence of topology and many-body localization in a driven system, the avoidance of heating, and the topologically-protected dynamical control of quantum states. A brief summary is given in Fig.~\ref{fig:triangle}.

\textit{Model.} We consider a one-dimensional spin-$\frac{1}{2}$ Hamiltonian consisting of a time-periodic and a time-independent contribution,
$
     H(t)=H_1(t)+H_2.
$
The time-periodic part is given by
\begin{align}
  H_1(t) &= \sqrt{\omega}\sin(\omega t) \sum_{i=1}^{L/2} \alpha_i \sigma^z_{2i-1} \sigma^z_{2i}\label{eq:H1_t}\\
  +
  \sqrt{\omega}&\cos(\omega t+\phi_0) \sum_{i=1}^{(L-1)/2} \left( \beta_i \sigma^y_{2i} \sigma^z_{2i+1} + \gamma_i \sigma^z_{2i} \sigma^y_{2i+1} \right),\nonumber
\end{align}
where $\omega$ denotes the driving frequency. The time-independent contribution reads
\begin{equation}\label{eq:H2}
  H_2 =  \sum_{i=1}^{L-1} V_{i} \sigma^{x}_{i} \sigma^{x}_{i+1} + \sum_{i=1}^{L} h_{i} \sigma^{x}_{i},
\end{equation}
with $\sigma_i^{x,y,z}$ being the Pauli matrices. The prefactors $h_i,V_i,\alpha_i,\beta_i,\gamma_i$ can be chosen differently for each lattice site $i$, and we draw them from a random Gaussian distribution with a standard deviation of $\sigma_{h,V,\alpha,\beta,\gamma}$ in order to drive the system into a many-body localized phase.  %For now, we set $\phi_0=0$, but we will relax this condition below.

\begin{figure}[t]
  \includegraphics[width=\linewidth,clip]{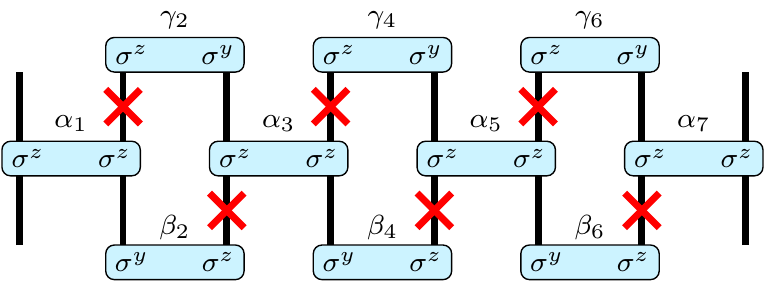}
  \caption{Representation of the non-vanishing part of the commutator $[H_1(t),H_1(t')]$ relevant to the second order Magnus expansion for a chain consisting of 8 sites. The vertical lines represent the sites, while the light blue rectangles represent the terms acting on the sites. The red crosses mark terms which are commuting even though they are acting on the same site. The commutator of two two-body interactions effectively generates a three-body interaction.}\label{fig:comm}
\end{figure}

The above Hamiltonian is time periodic, $H(t+T)=H(t)$, with $T=2\pi/\omega$
\changes{for all times $t$}. In order to gain some understanding of what physics one should expect to be modeled by $t\mapsto H(t)$, one can exploit the fact that the stroboscopic dynamics (which neglects the micro-motion) can be described by an effective \textit{time-independent} Hamiltonian $H^{\rm eff}$. In the high frequency limit $\omega\gg h_i,V_i,\alpha_i,\beta_i,\gamma_i$, this effective Hamiltonian can be obtained by virtue of a Magnus expansion \cite{magnus}, see supplemental information for details. An expansion up to order 
$\mathcal{O}\left({1}/{\sqrt{\omega}}\right)$ yields
 \begin{align}
    H^{\rm eff} &= \sum_{i=1}^{L-2} \lambda_{i} \sigma^z_i \sigma^x_{i+1} \sigma^z_{i+2} +\sum_{i=1}^{L-1} V_{i} \sigma^{x}_{i} \sigma^{x}_{i+1} + \sum_{i=1}^{L} h_{i} \sigma^{x}_{i} ,\label{eq:Heff}
\end{align}
with coupling terms
\begin{equation}\label{eq:efflambda}
    \lambda_{i,\textnormal{odd}} = \cos(\phi_0)\alpha_i \beta_{i} ,~~\lambda_{i,\textnormal{even}} =\cos(\phi_0) \gamma_i \alpha_{i+1}.
\end{equation}
The second and third term in Eq.~\eqref{eq:Heff} arise from a first order Magnus expansion (which amounts to a time average) of $H_2$ (the first order expansion of $H_1$ vanishes as the time-average is zero). The first term in Eq.~\eqref{eq:Heff}, which is an effective three-spin interaction, arises from the second order Magnus expansion which involves a time-averaged commutator of $[H_1(t),H_1(t')]$, see Fig.~\ref{fig:comm} for a pictorial representation. All of the other commutators in the second order Magnus expansion as well as all higher-order terms scale away with ${1}/{\sqrt{\omega}}$ or faster. We note that by this we have explicitly constructed an effective three-body Hamiltonian from a time-dependent two-body one. We choose to Floquet engineer this particular example of a three body Hamiltonian as it \changes{has been} shown before to harbor intriguing physics for which disorder coexists with \changes{symmetry protected} topological features \cite{topmbldyn_bahri,topmbldyn_yao,firstpaper,goihl2019}, which we will discuss in more detail in the results section.  However, one should note that this procedure is general and can be used to engineer effective three body terms of a different desired form in the same way as illustrated here.  

A benchmark of how well the Floquet-engineered physics of the time-dependent Hamiltonian given in Eqs.~(\ref{eq:H1_t}) and (\ref{eq:H2}) agrees with that of the effective time-independent Hamiltonian $H^{\rm eff}$ of Eq.~(\ref{eq:Heff}) is shown in Fig.\ \ref{fig:dis} as well as in the supplemental information. We find perfect agreement in the large-frequency limit where the Magnus expansion is justified. We emphasize that the results in this work are obtained using the full time-dependent Hamiltonian and the effective Hamiltonian $H^{\rm eff}$ only facilitates the physical interpretation.

\textit{Results.} The beauty of the effective Hamiltonian in Eq.~(\ref{eq:Heff}) is that it can host a symmetry protected topological phase with protected gapless spin-$\frac{1}{2}$ edge excitations that can be used to define a qubit \cite{topmbldyn_bahri,topmbldyn_yao}. %{\color{red} \st{If $V_i=0=h_i$, the edge excitation on the left edge is described by the edge operators $L_x=\sigma_{1}^{x}\sigma_{2}^{z}$, $L_y=\sigma_{1}^{y}\sigma_{2}^{z}$, $L_z=\sigma_{1}^{z}$. A similar spin-$\frac{1}{2}$ algebra can be constructed for the right edge} \cite{topmbldyn_bahri}.}
\changes{In the presence of disorder and interactions, many-body localization inhibits ergodicity and prevents the system from thermalizing, thereby extending the topological protection to high energies. The downside of this highly-desirable behavior is that Eq.~(\ref{eq:Heff}) contains three-body interactions that are generally not available. It is key to the understanding of the significance of the present scheme to acknowledge that no nearest neighbor Hamiltonian $H = \sum_i (A_i B_{i+1}+ \text{h.c.}) $ can give rise to the ground state of $H^{\rm eff}$ \cite{Nielsen}. The Floquet approach presented here hence overcomes a road block against preparing such states of matter, as native three-body interactions of this type are generally not available. Cluster states as ground states of $H^{\rm eff}$ can be approximated by nearest-neighbor Hamiltonians \cite{PhysRevA.74.040302,Gadgets},
but they require \changes{very strong interactions and high levels of 
control to effectively arrive at Hamiltonians with 
higher locality in perturbation theory}}.
%
% Sorry, ok!

Floquet engineering overcomes this hurdle \cite{PhysRevLett.119.123601} and moreover provides a natural way to dynamically control the edge qubit. Normally, a periodically-driven system would heat up and eventually approach an infinite-temperature state. In our case, however, many-body localization suppresses heating. In a nutshell, the combination of Floquet engineering, topology, and many-body localization can be utilized to level disadvantages, while harvesting the full advantages of the respective subfields. This allows one to implement and dynamically control a spin-$\frac{1}{2}$ qubit at high energies. We illustrate the three different aspects --- disorder, heating, and control --- separately in the following sections (see Fig.~\ref{fig:triangle} for a schematic summary). %We simulate the full time dependent Hamiltonian \je{$t\mapsto H(t)$} given in Eqs.~(\ref{eq:H1_t}) and (\ref{eq:H2}) using exact diagonalization up to $L=12$ lattice sites.

\begin{figure}[t]
  \includegraphics[width=\linewidth,clip]{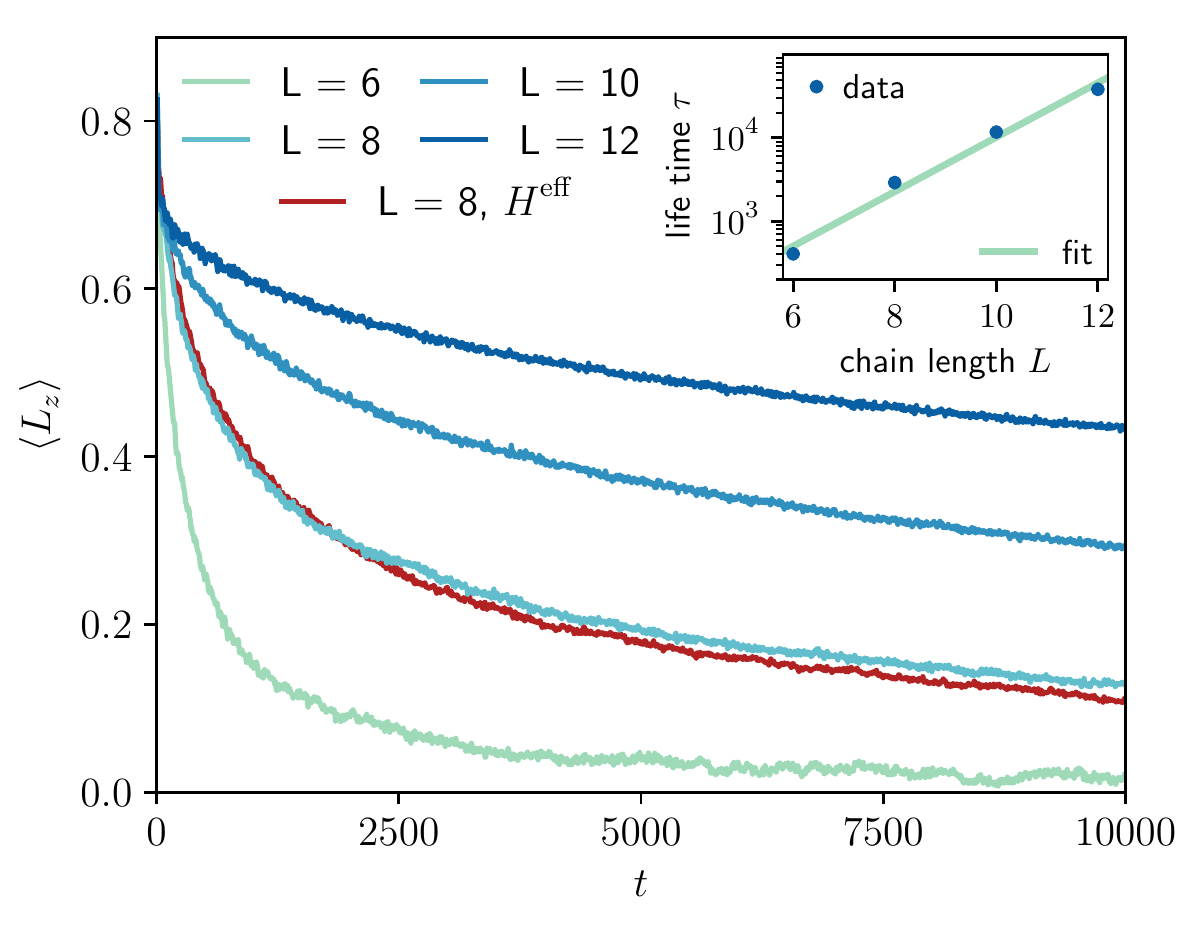}
  \caption{\textit{Topological protection at high energies.} Time evolution of the edge spin $\langle L_z\rangle$ governed by Eqs.~(\ref{eq:H1_t}) and (\ref{eq:H2}) for different system sizes and a driving frequency of $\omega = 1000$. \changes{The edge spin decays exponentially with $\sim {\exp} [-\sqrt{t/\tau}]$. As the system size increases, the life time  $\tau\sim \exp[0.752L]$ } of the spin becomes exponentially large in the system size (inset) due to the protection by topology and many-body localization even at high energies. The deviation from $\langle L_z\rangle=1$ is due to the fact that although this operator has a non-zero overlap to the operator describing the topologically-protected gapless edge mode they are not identical. We choose $\phi_0=0$, disorder strengths $\sigma_{\alpha,\beta,\gamma}= 1.0$, $\sigma_V = 0.1$, $\sigma_h=0.05$, and average over $1000$ random configurations. For $L=8$, we show data obtained using the effective time-independent Hamiltonian of Eq.\ (\ref{eq:Heff}) for comparison.}
  \label{fig:dis}
\end{figure}

\textit{Topological protection.} The time-dependent Hamiltonian in Eqs.~(\ref{eq:H1_t}) and (\ref{eq:H2}) hosts a \changes{symmetry protected} topological phase at high energies since the system is barred from thermalization by many-body localization (\changes{in the sense that
edge modes remain protected and suitable non-local order parameters take non-zero
values even away from ground states}). One can illustrate this explicitly by demonstrating that the gapless spin-$\frac{1}{2}$ edge mode has an infinitely long life. To this end, we prepare the system in a product state in which all the spins are initially pointing along the positive $z$-direction. The driving frequency is chosen as $\omega=1000$. We calculate the time evolution of the expectation value of the boundary spin  \changes{$L_z=\sigma_1^z$}  using the full $t\mapsto H(t)$ in Eqs.~(\ref{eq:H1_t}) and (\ref{eq:H2}). The operator describing the exact topologically-protected gapless spin-$\frac{1}{2}$ edge mode at the left boundary has a non-zero overlap with $L_z$ and is of the form $\tilde{L}_z = a\, L_z + b\, B_z$ \changes{for suitable complex $a$ and $b$} with $B_z$ being a bulk contribution that does not contribute in the thermodynamic limit. The results are summarized in Fig.~\ref{fig:dis}. As the system becomes larger, topological protection becomes increasingly robust, and the life time of $\langle L_z\rangle$ becomes exponentially large. We emphasize that the spins $\langle \sigma^z_i\rangle$ away from the edge decay \changes{quickly} for arbitrary $L$ since they do not have a finite overlap with a topologically-protected mode (data not shown). In a nutshell, Fig.~\ref{fig:dis} illustrates that topology and many-body localization team up so that information can be stored robustly at the edge of a driven system at high energies.

\begin{figure}[t]
\includegraphics[width=\linewidth,clip]{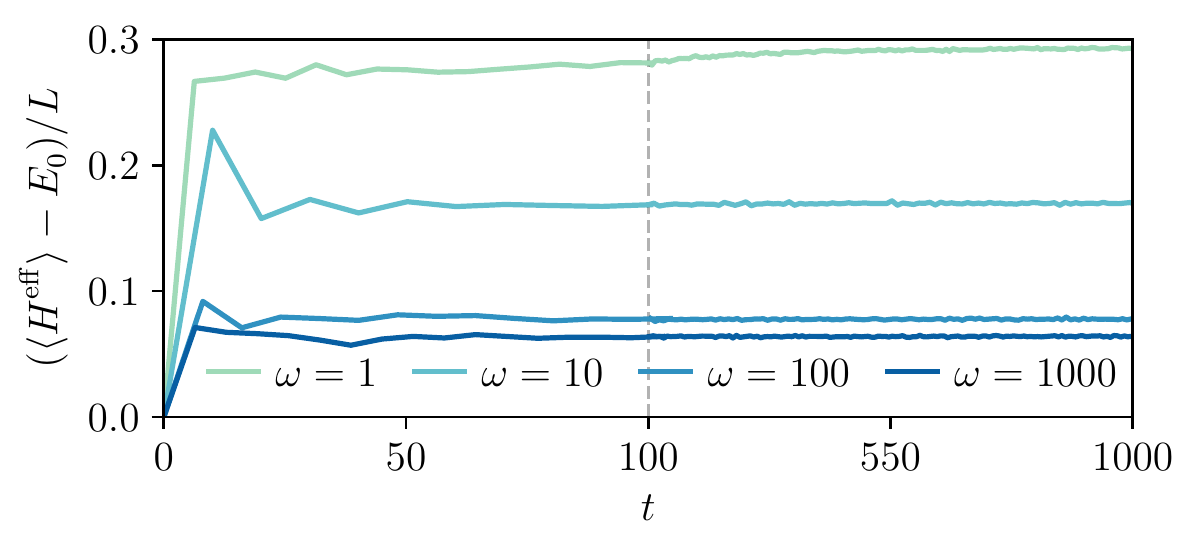}
\caption{\textit{Suppression of heating.} Time evolution of the excess energy pumped into the system by the periodic drive. The energy saturates quickly, and the system does not approach an infinite-temperature state. This suppression of heating is due to many body localization and detuning from single-particle resonances. We choose $L=10$, $\phi_0=0$, $\sigma_{\alpha,\beta,\gamma}=1.0$, $\sigma_V=0.1$, $\sigma_h=0.05$ and average over $1000$ random configurations. Note that the scaling of the linear time axis changes at $t=100$.}
%average $\sigma_V$ and $\sigma_h$ over $1000$ random configurations. Note that the scaling of the linear time axis changes at $t=100$.}
\label{fig:heating}
\end{figure}

\textit{Heating.} An interacting, periodically-driven quantum system is generically expected to heat up over time \cite{floquet_alessio}. This would be detrimental to our aim of storing information in the topologically-protected edge states, as the system would heat up to infinite temperature, an equal superposition of all many-body states with no memory of the initial state. However, many-body localization bars the system from thermalizing \cite{mbl_huse,mbl_rmp} \changes{and it has been shown that this can be exploited to push the effects of heating to exponentially large times \cite{noheating1,noheating2}, where the scaling of the heating is determined by the localization length in the system.} This is less obvious than it may seem, as many-body localized systems still allow for quantum information propagation \cite{Local}, a feature that is reflected by a logarithmic entanglement growth in time following global quenches \cite{Prosen_localisation,Pollmann_unbounded}. At the same time, an approximate emergent picture of $L$ quasi-local constants $C_j$, $j=1,\dots, L$, of motion emerges for $H^{\rm eff}$ that commute with each other 
and with $H^{\rm eff}$. This suppression of driving-induced heating by the many-body localization can be understood by acknowledging that many-body localized systems behave very similar to so-called integrable ones, which feature an extensive set of (quasi-)local constants of the motion. These integrable systems are known to feature interesting periodically driven long-term states immune to runaway heating in contrast to generic (non-integrable) driven systems \changes{\footnote{\changes{Outside of the many-body localized context we are discussing here exceptions to this rule where even integrable systems can heat up are known \cite{PhysRevLett.120.220602}}}}.

Detuning away from single particle resonances by going to high frequencies has a similar effect. Therefore, a system like ours is doubly protected from heating by employing both mechanisms: many-body localization and detuning. This is illustrated in Fig.~\ref{fig:heating}, where we monitor the time evolution of the energy pumped into the system by the periodic drive in $t\mapsto H(t)$. We initially prepare the system in the ground state of the effective Hamiltonian in Eq.~\eqref{eq:Heff} of the corresponding Magnus expansion. The energy initially increases as the drive produces excess energy, which, if redistributed thermally, would translate to heating. As we approach the high-frequency limit and Floquet-engineer the many-body localized Hamiltonian in Eq.~\eqref{eq:Heff}, excess heat production ceases and the system remains close to its ground state for arbitrary large times. This demonstrates that many-body localization and Floquet engineering conspire to efficiently suppress heating, which would be detrimental to storing information.

\begin{figure}[t!]
\includegraphics[width=\linewidth,clip]{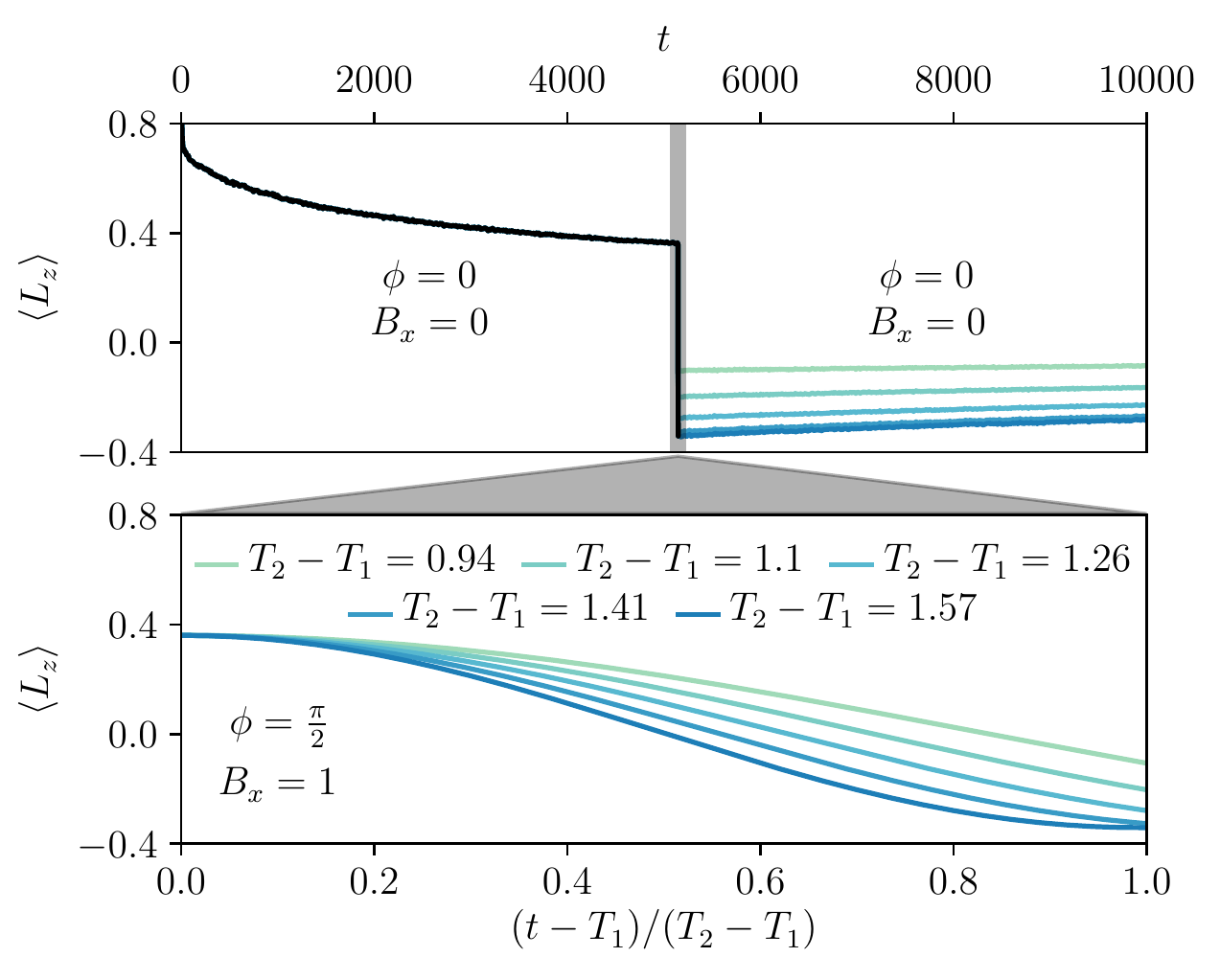}
\caption{\textit{Dynamical control.} The edge spin can be flipped on a short time scale $T_1 \le t \le T_2$, $T_2-T_1\sim 1$ by applying a constant magnetic field $B_x$ as well as a brief $\pi/2$ phase shift to the drive (lower panel). Before ($t<T_1$) and after ($t>T_2$), the spin is stable on exponentially-long time scales (upper panel). We choose $L=10$, a driving frequency of $\omega=1000$, and disorder strengths $\sigma_{\alpha,\beta,\gamma}=1.0$, $\sigma_V=0.1$, $\sigma_h=0.05$. Disorder averages are performed over $1265$ random configurations.} %. Amount of periods needed to flip spin is $4h/\omega=250$.}
\label{fig:control}
\end{figure}

\textit{Control.} Finally, we demonstrate how to dynamically control the Floquet-engineered system. Our aim is to manipulate the edge spin on a short time scale by changing the parameters of the drive. To this end, we introduce a time-dependent phase $t\mapsto \phi(t)$ replacing $\phi_0$ in Eq.~\eqref{eq:H1_t}, which in general breaks the time-periodicity. However, in either the case of (i)  a slowly varying $\phi$, or (ii) a piece-wise constant phase $\phi$, the above arguments still hold within each time-interval in which $\phi$ is (approximately) constant. Changing the phase of the drive allows us to effectively control the system as the phase governs the magnitude of the couplings $\lambda_i$ in the effective Hamiltonian Eq.~\eqref{eq:Heff} via Eq.~(\ref{eq:efflambda}). 
We will now concentrate on the following, particularly instructive, example
\begin{equation}
  \phi(t)=\begin{cases}0 & t <T_1 \;\;\;\; {\rm{or}} \;\;\;\; t>T_2,\\
  \pi/2 &T_1\leq t \leq T_2.
  \end{cases}
\end{equation}
We aim at rotating the spin and thus for times $T_1\leq t\leq T_2$ add a constant field along the $x$-direction, $H_2\mapsto H_2+B_x {\changes{\sum_{i} \sigma^{x}_{i}}}$. The effective (now piece-wise time-dependent) Hamiltonian from a second order Magnus expansion reads
\begin{align}
  H^{\rm eff}(t) =& \sum_{i=1}^{L-2} \lambda_{i}(t) \sigma^z_i \sigma^x_{i+1} \sigma^z_{i+2}+\sum_{i=1}^{L-1} V_{i} \sigma^{x}_{i} \sigma^{x}_{i+1}\notag\\&+ \sum_{i=1}^{L}\left(B_x(t)+ h_{i}\right) \sigma^{x}_{i} ,\label{eq:Heff_t}
\end{align}
where 
\begin{equation}
  \lambda_\text{i,odd\;(even)}(t)=\begin{cases}  \alpha_i \beta_{i}\; (\gamma_i \alpha_{i+1}) & t <T_1  ~{\rm{or}}~ t>T_2\\
  0& T_1\leq t \leq T_2.
  \end{cases}
\end{equation}
and $B_x(t)=B_x$ for $T_1 \leq t \leq T_2$.

In Fig.~\ref{fig:control}, we present results for the time evolution of the edge spin $\langle L_z\rangle$ for different switching times $T_2-T_1\sim 1$, $B_x=1.0$, and a driving frequency of $\omega=1000$ (disorder parameters are given in the caption). We again stress that the time-evolution is calculated using the full time-dependent Hamiltonian $t\mapsto H(t)$ and that its high-frequency counterpart, the effective Hamiltonian $H^{\rm eff}$, only facilitates the interpretation of the results. For times $t<T_1$, the edge state is protected by topology, and its information (i.e., being in the up-state) can be stored for exponentially long times. At $t=T_1$, the phase of the driven Hamiltonian is switched from $\phi_0=0$ to $\pi/2$, and a magnetic field is applied. The magnetic field flips the direction of the spin.  At time $t=T_2$, both the phase and the magnetic field are switched back off, $\phi_0=B_x=0$. The new state (where the edge spin now points down) can again be stored for exponentially long times as the topological properties are restored. This exemplifies how the versatility inherent to Floquet engineering can be utilized to dynamically control and manipulate the information in the topologically-protected edge states.

\textit{Conclusion.} We have demonstrated how quantum states can be dynamically manipulated in a stable way in a periodically-driven, realistic spin system. Starting from a simple spin model with two-body interactions, Floquet-engineering allows one to design a toy Hamiltonian in which topology and disorder conspire to  protect (qubit) spin states even away from low energies as suggested in Refs.~\cite{topmbldyn_bahri,topmbldyn_yao}. We explicitly demonstrated how heating is suppressed efficiently and investigated how well the toy Hamiltonian approximates the dynamics governed by the original one. Taking these properties together, this could provide a fascinating new route to implementing and controlling stable qubits even at high temperatures, which should be the subject of future research.

\textit{Acknowledgments.} This work has been supported by the DFG through the CRC 183 (Projects A01, A03, and B01), the DFG FOR 2724,  the Cluster of Excellence Matter and Light for Quantum Computing (ML4Q) EXC 2004/1 - 390534769, and through the Emmy Noether program (KA 3360/2-2). This  work  has  also received  funding  from  the  European  Union's  Horizon2020  research  and innovation  programme  under  grant  agreement  No.~817482 (PASQuanS). We further acknowledge support from the 
Max Planck--New York City Center for Non-Equilibrium Quantum Phenomena.

\end{document}